\documentclass[aps,prb,twocolumn,floatfix,amsmath,amssymb,superscriptaddress,]{revtex4-2}
\usepackage[final]{graphicx}
\usepackage{bm}
\usepackage{afterpage}
\usepackage{color}
\usepackage{comment}
\usepackage[colorlinks=true,urlcolor=blue,linkcolor=blue,citecolor=blue]{hyperref}
\usepackage{booktabs}
\usepackage{siunitx}
\usepackage{braket}
\usepackage{gensymb}
\usepackage[version=4]{mhchem}

\usepackage{afterpage}

\begin{document}
\title{Enhanced anomalous Hall conductivity via Ga doping in Mn\textsubscript{3}Sn and Mn\textsubscript{3}Ge}

% x ,<--> 1-x
% 比较
% 掺杂实验

\author{Chenyue Wen}
\thanks{These authors contributed equally.}
\affiliation{Fert Beijing Institute, MIIT Key Laboratory of Spintronics, School of Integrated Circuit Science and Engineering, Beihang University, Beijing 100191, China}
\affiliation{Integrated Circuit and Intelligent Instruments  Innovation Center, Qingdao Research Institute of Beihang Universit, Qingdao 266100, China}
%\affiliation{National Key Laboratory of Spintronics, Hangzhou International Innovation Institute, Beihang University, Hangzhou, 311115, China}
\affiliation{State Key Laboratory of Spintronics, Hangzhou International Innovation Institute, Beihang University, Hangzhou 311115, China }
\author{Danrong Xiong}
\thanks{These authors contributed equally.}
\affiliation{Truth Memory Corporation, Beijing 100088, China}
\author{Chengyi Yang}
\affiliation{Fert Beijing Institute, MIIT Key Laboratory of Spintronics, School of Integrated Circuit Science and Engineering, Beihang University, Beijing 100191, China}
\affiliation{Integrated Circuit and Intelligent Instruments  Innovation Center, Qingdao Research Institute of Beihang Universit, Qingdao 266100, China}
\affiliation{State Key Laboratory of Spintronics, Hangzhou International Innovation Institute, Beihang University, Hangzhou 311115, China }
\author{Dapeng Zhu}
\email{zhudp@buaa.edu.cn}
\affiliation{Integrated Circuit and Intelligent Instruments  Innovation Center, Qingdao Research Institute of Beihang Universit, Qingdao 266100, China}
\author{Weisheng Zhao}
\email{weisheng.zhao@buaa.edu.cn}
\affiliation{Fert Beijing Institute, MIIT Key Laboratory of Spintronics, School of Integrated Circuit Science and Engineering, Beihang University, Beijing 100191, China}
\affiliation{Integrated Circuit and Intelligent Instruments  Innovation Center, Qingdao Research Institute of Beihang Universit, Qingdao 266100, China}
\affiliation{State Key Laboratory of Spintronics, Hangzhou International Innovation Institute, Beihang University, Hangzhou 311115, China }

\begin{abstract}
This study examines the anomalous Hall effect (AHE) in the Heusler series  \ce{Mn3Z}  (Z=Ga, Ge, Sn), with a particular emphasis on the manipulation of non-collinear antiferromagnetic structures to enhance AHE. By employing density-functional theory and first-principles calculations, we demonstrate that the anomalous Hall conductivity  is markedly responsive to electron filling. By strategically doping Ga into \ce{Mn3Sn} and \ce{Mn3Ge} in order to modulate the electron density, a significant increase in anomalous Hall conductivity (AHC) is achieved. It is noteworthy that a Ga:Sn ratio of 1:5 yields peak AHC values exceeding $\mathrm{700(\Omega \cdot cm)^{-1}}$, while 3:7 Ga-Ge ratios can result in AHC values surpassing $600\mathrm{(\Omega \cdot cm)^{-1}}$. A comparison between the virtual crystal approximation and supercell construction methods for doping has revealed consistent trends. The results of this study pave the way for optimizing AHE in non-collinear AFM materials.  %. 指代non-colinear
%, which could potentially lead to improvements in antiferromagnet-based technologies for switching and storage applications.% 修改
\end{abstract}

\maketitle
\section{Introduction}
The progress in big data, cloud computing, and artificial intelligence has driven the demand for innovative memory technologies that can adapt to increasing density and reduced power usage while sustaining high-speed and non-volatile capabilities~\cite{cui2024prospects}.
Compared to ferromagnetic(FM) devices, antiferromagnetic (AFM) devices have risen to prominence in the realm of ultrafast, ultrahigh-density storage, owing to their unique benefits:  no stray fields; high resonance frequency; and robustness against external magnetic disturbances~\cite{gomonay2014spintronics,rimmler2024non,xiong2022antiferromagnetic}. Given their distinctive characteristics, such as zero net magnetic moment and immunity to external magnetic fields, the manipulation and detection of antiferromagnetic moments remains a formidable challenge,  representing a critical frontier in spintronics research~\cite{qin2020noncollinear,rimmler2024non,yan2020electric}. To decipher the magnetic structure of AFM materials in device applications, a suite of detection techniques have been recently proposed. These include tunneling magnetoresistance~\cite{dieny2020opportunities}, AFM proximity effect~\cite{li2016antiferromagnetic}, exchange bias~\cite{du2023electrical}, and the anomalous Hall effect (AHE)~\cite{chen2014anomalous}. Among these, the AHE stands out as a particularly convenient method for detecting the magnetic structures of AFM materials, as it facilitates the development of pure antiferromagnetic devices. This is achieved without the need to engineer complex heterostructures such as  AFM/NM/AFM magnetic tunnel junctions (MTJs)~\cite{qin2023room}. Despite the advantages of using AHE, a significant challenge remains in achieving a robust AHE response. Consequently, it is of paramount importance to develop a strategy for generating substantial anomalous Hall conductivity (AHC) from antiferromagnetic materials.

\begin{figure}[htbp]
  \centering
  \includegraphics[width=8cm]{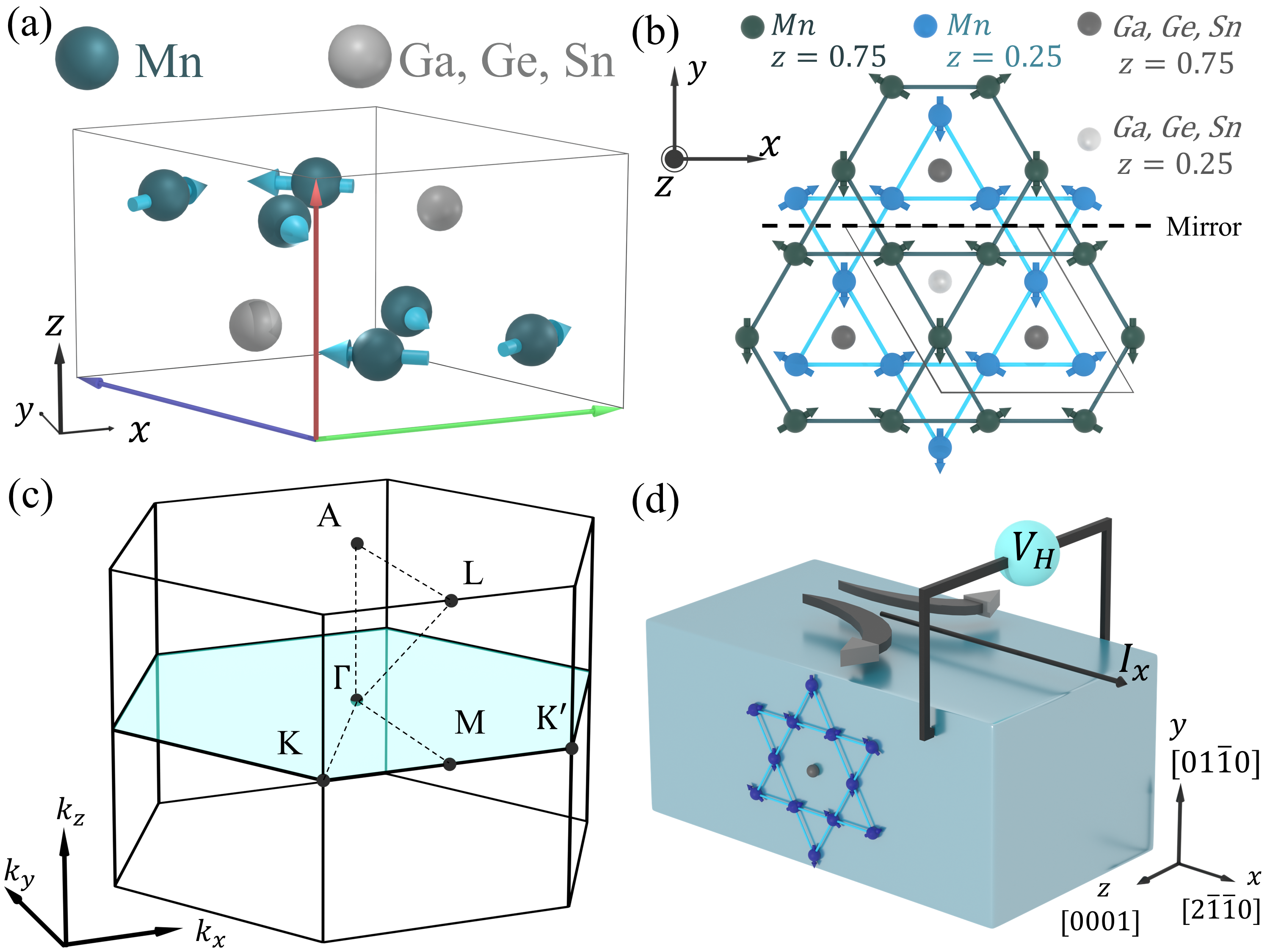}
  \caption{
    (a) The primitive cell of hexagonal \ce{Mn3Z} lattice. Magnetic moments of manganese atoms are marked by arrows. 
    (b) Top view of the lattice structure and magnetic structure of Mn\textsubscript{3}Z, showing the $\Gamma_{1g}(A_y)$ magnetic structure. The gray dashed lines indicate the $M_y$ mirror plane of the symmetry operation $\{ M_y| \tau = c/2 \}$. 
    (c) The first Brillouin zone and its high symmetry points. 
    (d) Schematic diagram showing the orientation of AHE and kagome surfaces.
  }
  \label{fig1}
\end{figure}

The intrinsic AHE is conventionally believed to be proportional to the magnetization in ferromagnetic materials. Nonetheless, in a select group of non-collinear antiferromagnets, a non-negligible AHE is observed even in the absence of a net magnetic moment. A prime example of such materials is the hexagonal \ce{Mn3Z} (where Z can be Ga, Ge, or Sn), which has become the focus of intensive research~\cite{xu2023robust,wang2021robust,song2024large,wang2022noncollinear}. This compound exhibits a non-collinear antiferromagnetic order at a 120° angle~\cite{brown1990determination,soh2020ground}, characterized by a unique Kagome lattice structure formed by Mn atoms within the (0001) crystallographic plane, as depicted in Figs. \ref{fig1}(a) and (b).

% 实验可行性
This arrangement that derived from geometric frustration of exchange interactions on the kagome lattice of Mn atoms, disrupts time-reversal symmetry. In conjunction with spin-orbit coupling (SOC), it leads to a non-zero, non-odd parity Berry curvature (BC) in reciprocal space~\cite{yang2017topological}, which explains the observation of significant anomalous Hall conductivity in recent research. In these studies, the anomalous Hall conductivity (AHC) of \ce{Mn3Sn} was observed to be up to 100 $\mathrm{(\Omega \cdot cm)^{-1}}$ at 100 K ~\cite{nakatsuji2015large}, while the AHC of \ce{Mn3Ge} can reach 300 $\mathrm{(\Omega \cdot cm)^{-1}}$ at 2 K~\cite{nayak2016large,chen2021anomalous}.  These values are comparable to some ferromagnets~\cite{nagaosa2010anomalous} and consistent with recent first-principles calculations ~\cite{zhang2017strong,guo2019erratum,li2023field}. These calculations, together with experiments~\cite{yang2017topological,kubler2018weyl,kuroda2017evidence}, also demonstrated their Weyl semimetallic nature. 

These calculations reveal a striking sensitivity of the AHC to fluctuations in the Fermi level ($E_f$). Notably, in both \ce{Mn3Sn} and \ce{Mn3Ge}, a significant peak emerges in the energy-dependent AHC curves, positioned approximately $0.11\ \text{eV}$ below the $E_f$. This observation suggests that enhancing the AHC may be feasible through hole doping, specifically by substituting some Sn or Ge with Ga. The rationale behind this approach is that \ce{Mn3Ga} experiences a downward shift in the Fermi level by about $0.34\ \text{eV}$, attributable to gallium (a group III element) possessing one fewer valence electron compared to germanium or stannum (group IV elements). Additionally, the non-magnetic nature of gallium ensures that it minimally disrupts the configuration and magnetic properties of the Mn atoms within the kagome lattice.

In this work, we conduct an ab-initio investigation into the AHE in Ga-doped hexagonal \ce{Mn3Sn} and \ce{Mn3Ge}. The structure of this paper is as follows: Section II delineates our methodology, which is grounded in density-functional theory (DFT) and ab initio Berry curvature calculations. We detail the computational parameters and approaches employed to determine the AHC. Next section juxtaposes two distinct computational Ga-doping methods to facilitate a comparative analysis. Then we present our key results, highlighting the doping-dependent AHC curves that emerge from our calculations. Concluding remarks and a summary of our findings are provided in Section IV.

\section{method and computational details}
 %In this study, We construct ordered $Mn_3Z$ lattice with experimental lattice constants (as presented in Table 1) and the ideal Wyckoff position of the Mn $6h$ atomic sites ($x=5/6$), which represents a scenario without trimerization(its impact will be discussed in appendix). In the case of a doped lattice, Vegard's law~\cite{vegard1921konstitution} is employed, whereby the new lattice constants are calculated as a weighted average of the two original lattice constants. The initial magnetic order was set to be $\Gamma_{1g}(A_y)$(as indicated in [Fig.~\ref{fig1}(b)]), where Mn moments align inside the $x-y$ plane and form $120\degree$ angles with neighboring moment vectors. 
 
 In this study, we constructed an ordered \ce{Mn3Z} lattice utilizing the experimental lattice constants as detailed in Table \ref{tab1}, along with the ideal Wyckoff position for the Mn $6h$ atomic sites at $x = 5/6$. This configuration corresponds to a scenario devoid of trimerization, the implications of which are further explored in the appendix \ref{appE}. For the doped lattice, we applied Vegard's law~\cite{vegard1921konstitution} to calculate the new lattice constants as a weighted average of the original pair. The initial magnetic order was established as $\Gamma_{1g}(A_y)$, as illustrated in Figure \ref{fig1}(b), wherein the Mn moments are aligned within the x-y plane and form $120^\circ$ angles with adjacent moment vectors.

 The electronic ground states are calculated using density-functional theory with the QUANTUM ESPRESSO~\cite{giannozzi2009quantum} package. We incorporated SOC  into our calculations by utilizing fully relativistic ultra-soft pseudo-potentials from the PSLIBRARY~\cite{dal2014pseudopotentials}. For comparison, calculations without SOC were performed using scalar relativistic pseudo-potentials, which resulted in a strictly zero AHC in all subsequent calculations. The exchange-correlation potential was approximated using the Perdew-Burke-Ernzerhof (PBE) functional within the generalized gradient approximation (GGA) framework~\cite{perdew1996generalized}.  For the self-consistent  calculations, a $\Gamma$-centered k-point mesh of $8 \times 8 \times 9$ was utilized, with a plane-wave cutoff energy set at 70 Ry.
 
\begin{table}[h]
\centering
\caption{Experimental lattice constants (\si{\angstrom}) for \ce{Mn3Z}}
\begin{tabular}{cccc}
	\hline
	 $\indent$& \ce{Mn3Sn}~\cite{zimmer1972investigation} & \ce{Mn3Ge}~\cite{kadar1971neutron} &  \ce{Mn3Ga}~\cite{kren1970neutron}  \\
	 \hline
	 a&5.665&5.360&5.400\\
	 c&4.531&4.320&4.353\\
	 \hline
\end{tabular}
\label{tab1}
\end{table}

The anomalous Hall conductivity is calculated using the Wannier interpolation approach~\cite{wang2006ab}. Following the non-self-consistent calculation of 220 Bloch wave functions at each k-point, the Wannier90~\cite{mostofi2014updated} code is employed to construct the maximally-localised Wannier functions (MLWFs). At this juncture, we elected to utilise 68 Wannier functions per unit cell, encompassing the 3d, 4s orbitals of Mn and the 4s, 4p(5s, 5p)  orbitals of Ge and Ga(Sn) as preliminary projections, while excluding the semi-core orbitals. Subsequently, the anomalous Hall conductivity can be calculated using the WannierBerri~\cite{tsirkin2021high} package in accordance with the Kubo formula~\cite{xiao2010berry,gradhand2012first}:
\begin{align}
    \sigma_{\alpha \beta} = -\frac{e^2}{\hbar} \sum_n \int_{1BZ}{\frac{d^3\vec{k}}{(2\pi)^3} f_n(\vec{k})\Omega_{n}^{\alpha\beta}(\vec{k}) }
	\label{eq1}
\end{align}
where:
\begin{align}
	\Omega_{n}^{\alpha\beta}(\vec{k}) = -2\mathrm{Im} \sum_{m\neq n}\frac{ \bra{u_n(\vec{k})} \hat{v}_\alpha \ket{u_m(\vec{k})} \bra{u_m(\vec{k})} \hat{v}_\beta \ket{u_n(\vec{k})}}{(E_n(\vec{k})-E_m(\vec{k}))^2}
	\label{eq2}
\end{align}
is the Berry curvature in reciprocal space. The velocity operator is defined here as $\hat{v}=\frac{i}{\hbar}[\hat{H},\hat{r}]$ and $\alpha, \beta, \gamma$ represent the three Cartesian coordinates. $f_n(\vec{k})$ is the Fermi-Dirac distribution. At a temperature of 0 K, AHC $\sigma_{\alpha \beta}$ can be evaluated by summing all Berry curvature over the 1st Brillouin zone (1BZ) below the Fermi level. This integration over the 1BZ was conducted in a highly dense grid of $200 \times 200\times225$ points. To further enhance the precision of our calculations, we employed recursive adaptive refinement, conducting over 50 iterations.
% Recursive adaptive refinement of over 50 iterations is used for enhanced accuracy.

In order to mimic Ga-doping, we consider a $1\times 1 \times 2$ or $1\times 1 \times 3$ supercells containing 12 and 18 Mn atoms, and 4 and 6 Ge/Sn atoms, respectively. then replace some of them by Ga atoms. An alternative approach is to utilise the virtual crystal approximation (VCA) technique~\cite{bellaiche2000virtual}, which compositionally averaging the potentials of  Sn(Ge) and Ga: $V_{VCA} = xV_{Sn(Ge)}+(1-x)V_{Ga}$. However, in the case of half-doping, the original primitive cell is sufficient, and this will be discussed in greater detail in the subsequent section.

\section{AHC results and discussion}
\subsection{Half-doping: VCA vs non-VCA}
\label{sec3A}
\begin{figure}[htbp]
	\centering \includegraphics[width=8.5cm]{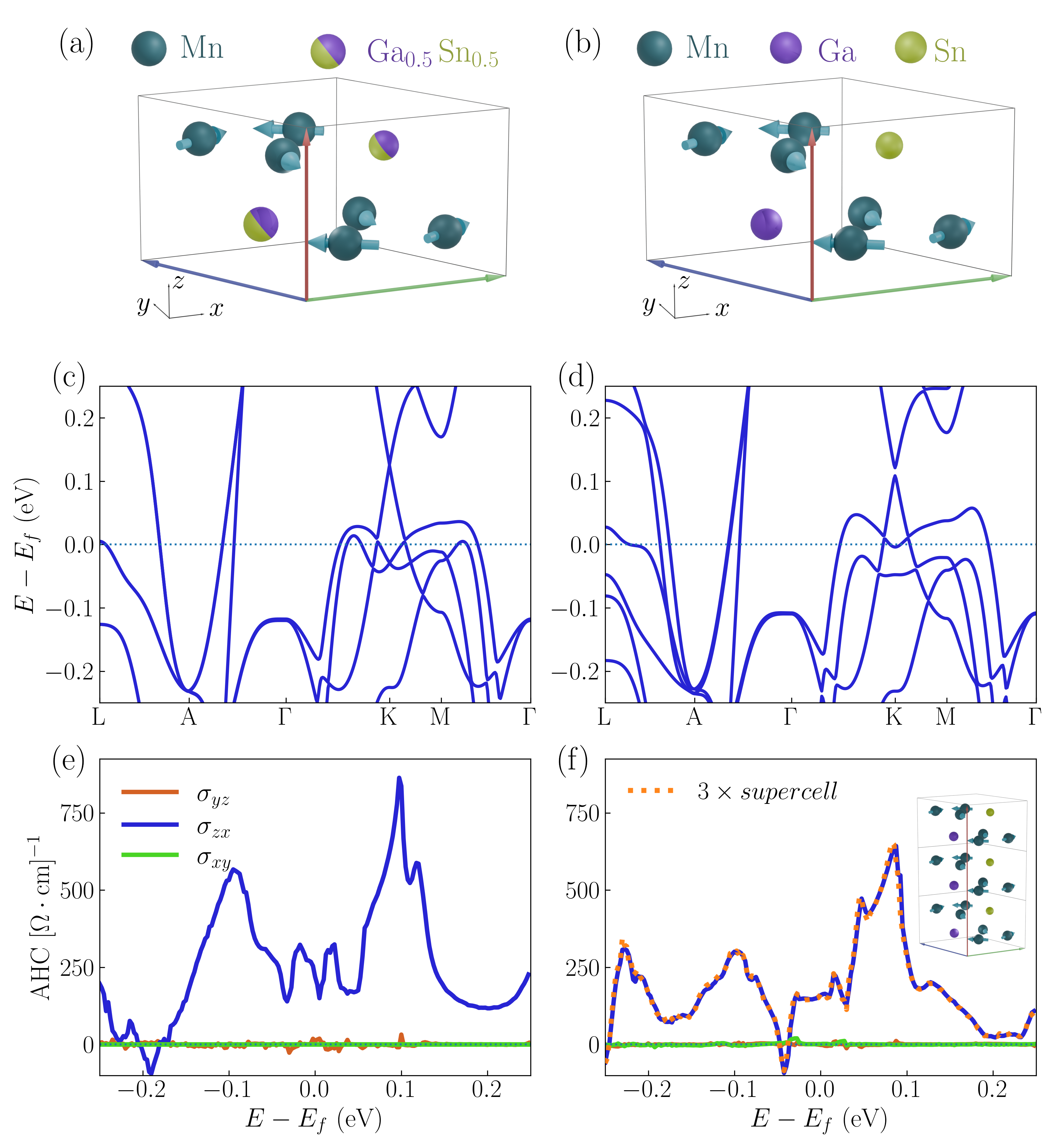} \caption{
		Validation of 2 distinct methods of Ga-doping. the primitive cells of the VCA approach (a) and the non-VCA approach (b) for DFT calculation. 
		The electronic band structure for the VCA (c) and non-VCA (d) scenarios and their energy-dependent AHC (e) and (f). The red dashed line in (f) represents a $1 \times 1\times 3$ supercell calculation,and the inset shows a schematic diagram of its primitive cell.
	}
	\label{fig2}
\end{figure}
Firstly, The two aforementioned doping approaches are validated. To elucidate, we examine the case of Mn\textsubscript{3}Ga\textsubscript{0.5}Sn\textsubscript{0.5}  which involves two distinct primitive cells as illustrated in Figs. \ref{fig2}(a) and (b). These figures correspond to the virtual crystal approximation (VCA) and non-VCA calculations, respectively. The band structures of these compounds are analyzed at high-symmetry points, indicated in Fig. \ref{fig1}(c). The corresponding band structure images are displayed in Fig. \ref{fig2}(c) and (d). The striking similarity between the two calculations is a consequence of their shared magnetic atom(Mn), magnetic structure, and valence electron count. Conversely, there is a notable discrepancy in the final AHC values, as depicted in Figs. \ref{fig2}(e) and (f). For the VCA case, the AHC along the x-z direction, is $159.45 \mathrm{(\Omega \cdot \text{cm})^{-1}}$, whereas for the non-VCA case, it is $225.57 \mathrm{(\Omega \cdot \text{cm})^{-1}}$. This variance can be attributed to the larger atomic radius and stronger SOC of Sn compared to Ga, which  can significantly influence its topological properties. The virtual crystal approximation (VCA) mitigates this difference through its compositional averaging of potentials, thereby diminishing the localized effects arising from the distinct nonmagnetic atoms.~\cite{eckhardt2014indirect}. Although the values differ, the overall trend remains consistent, with both exhibiting a peak approximately 0.1 eV above the Fermi level.

 The AHC in the other two directions is approximately 0, which is consistent with  the symmetry analysis~\cite{yang2017topological,zhang2017strong}: a mirror operation $\hat{M}_y$ reverses the sign of $\Omega_{x}(\vec{k})$ and $\Omega_{z}(\vec{k})$, thereby rendering their integration equal to zero. In contrast, $\Omega_{y}(\vec{k})$ remains unchanged, and thus retains a finite value. The small non-zero values of $\sigma_{yz}$ and $\sigma_{xy}$ are primarily due to the fact that the symmetry is not rigorously preserved during the Wannier interpolation step~\cite{sakuma2013symmetry}. Another magnetic structure, which is subject to a mirror operation with respect to the x-axis, will be discussed in Appendix \ref{appH}. Moreover, The results of $1\times 1\times 3$ supercell calculation are plotted in Fig. \ref{fig3}(f) with a red dashed line, and the schematic diagram of its primitive cell is  presented in the inset.  Upon repeating the primitive cell three times along the z direction, the results remained identical.

\subsection{Berry curvature distribution}

\begin{figure}[htbp]
	\centering \includegraphics[width=8.8cm]{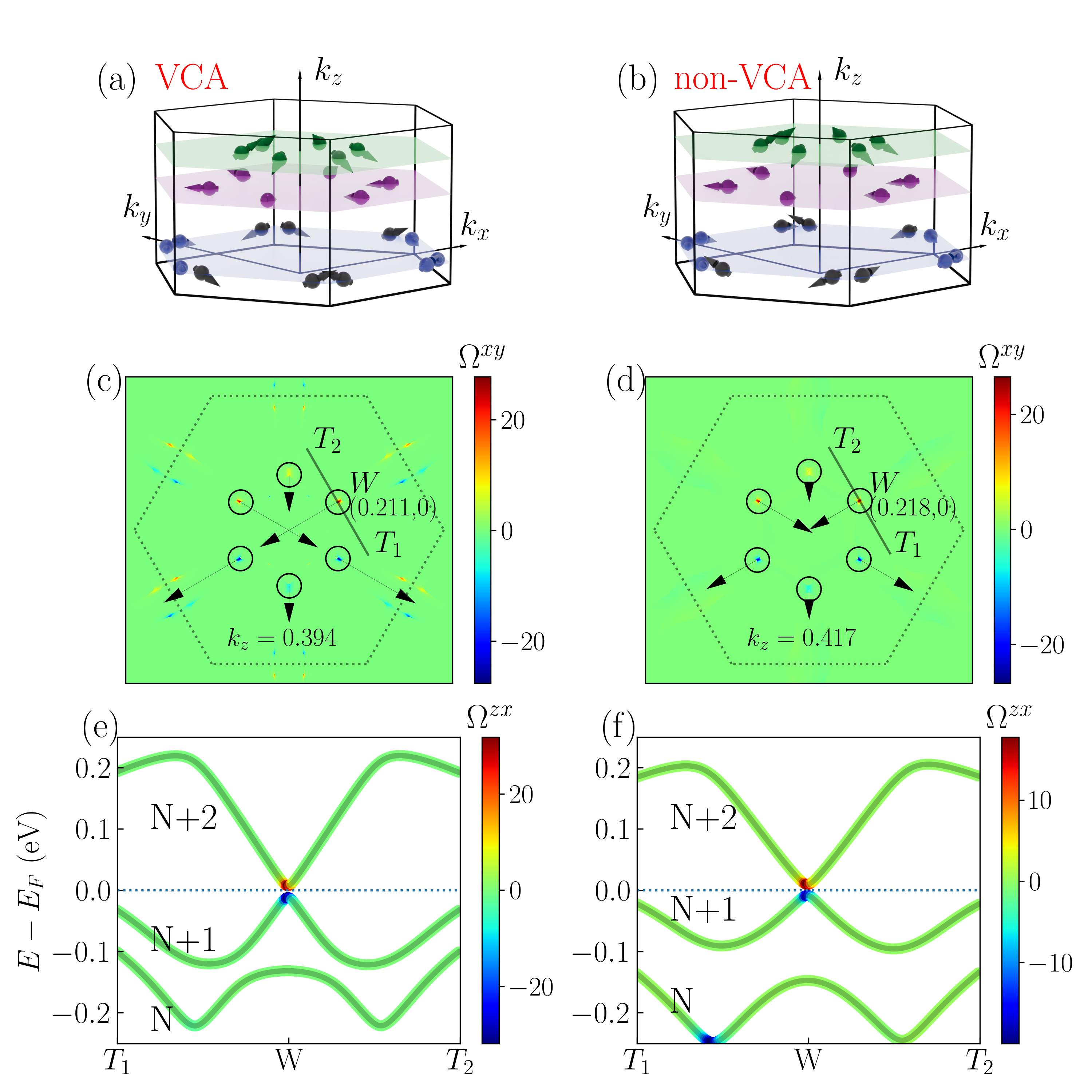} \caption{
		(a), (b): distributions of Berry curvature hotspots. both for the VCA approach and the non-VCA approach, respectively. Black arrows indicate BC hotspots with different directions. (c), (d): their top layers(green planes) of Berry curvature hotspots. The z component of the Berry curvature hotspots are represented by a pseudo-color image, while their x-y components are illustrated by the length and direction of the black arrows. Arrows of both images employ the same scale. The coordinates, which are expressed in units of reciprocal lattice vectors, of hotspots W are also marked on the plots. (e), (f): $N_v^{th}$, $(N_v+1)^{th}$, and$(N_v+2)^{th}$ Berry curvature fatbands in the vicinity of the hotspots. the k-path(solid black lines of (c)and (d)) is perpendicular to the orientation of their Berry curvature. The coordinates are expressed in units of reciprocal lattice vectors.
	}
	\label{fig3}
\end{figure}
Since the intrinsic AHE arises from the Berry curvature in reciprocal space, we performed a comprehensive analysis of the Berry curvature distribution to clarify the similarities and differences between the two methods. The Berry curvature is summed over all occupied bands and subsequently tabulated in the entire k-grid to identify the hot spots that contribute the most to the AHC.

The locations and orientations of the Berry curvature hotspots for both compounds in the upper portion of the 1BZ are illustrated in Figs.~\ref{fig3}(a) and (b).  The plots reveal striking similarities, with the position distributions of the hotspots for both VCA and non-VCA calculations differing by no more than $2.5\%$. Notably, aside from the six hotspots in the blue layer (indicated by black arrows), the orientation of the remaining Berry curvature hotspots (indicated by colored arrows) are also almost identical. Figs.~\ref{fig3}(c) and (d) present a detailed results of the green planes($k_z=0.394$ and $k_z=0.417$, in units of reciprocal lattice vectors), as their BC on the x-y plane indicated by a black arrows. Fig.~\ref{fig3}(c) provides a more detailed illustration of the preceding symmetry analysis: $ \Omega^{yz}(\vec{k})$ is an odd function of both $k_y$ and $k_x$, whereas $\Omega^{zx}(\vec{k})$ is an even function. In contrast, in the case of non-VCA, the occupation of the $2c$ site by different atoms results in the disruption of this symmetry. The small perturbation is reflected in the BC distribution in the reciprocal space (see Fig.~\ref{fig3}(d)). 
Subsequently, the Berry curvature fatbands in the vicinity of the hotspots were calculated. As illustrated in Fig.~\ref{fig3} (e) and (f), at both BC hotspots, linear dispersion relations featuring crossing points %akin to Dirac cones 
are observed between the $(N_v+1)^{th}$ and $(N_v+2)^{th}$ energy bands, where $N_v$ represents the total valence electron number. 
%This suggests that they may serve as potential candidates for Weyl points~\cite{jia2016weyl,chen2021anomalous,yang2017topological}. 
As illustrated in the fatbands, the entanglement between occupied and unoccupied states is markedly pronounced at these two points, contributing to a substantial neat Berry curvature, which is in accordance with the predictions of Eq.~\ref{eq2}. The energy gap at these points can be as small as 0.018eV, providing an explanation for the significant variation in the AHC with respect to the Fermi level. Once the electron filling deviates from this energy range, the contribution to the total AHC of these hotspots becomes zero. 
\subsection{Doping results}
\begin{figure}[htbp]
	\centering \includegraphics[width=8.6cm]{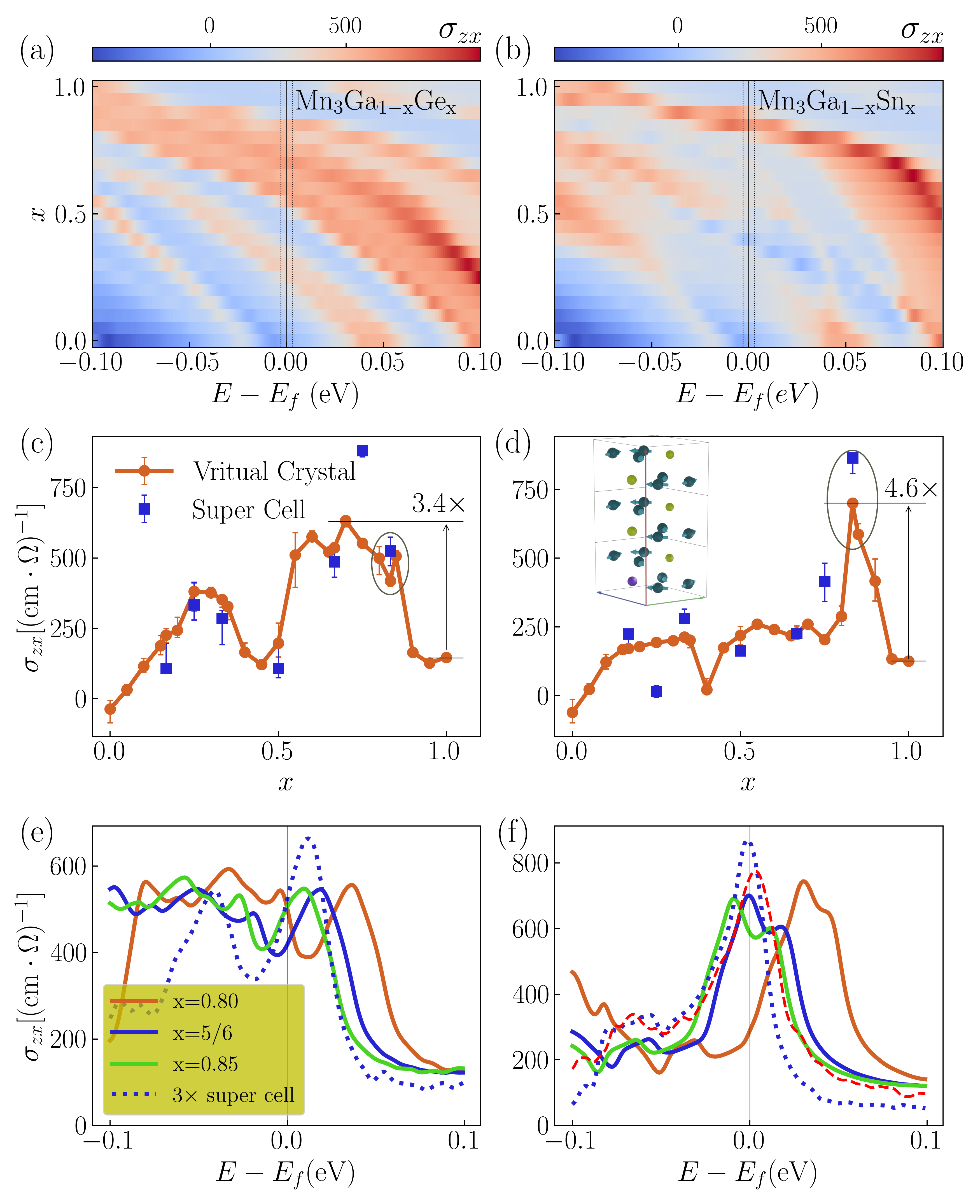} \caption{
		(a), (b): Energy- and doping-ratio-dependent AHC $\sigma_{zx}$ pseudocolor plot of Mn\textsubscript{3}Ga\textsubscript{1-x}Ge\textsubscript{x} and Mn\textsubscript{3}Ga\textsubscript{1-x}Sn\textsubscript{x}, respectively. The solid black line represents the neutral filling (Fermi level), while the dashed black line is employed to ascertain the error bar, which differs by 0.003 eV from the Fermi level. (c), (d): AHC that undergoes a change in response to alterations in the doping ratio for both VCA approach and supercell approach. The error bars represent the maximum and minimum values within the black dashed lines depicted in Figs (a) and (b) above. Inset shows a $1\times1\times3$ super cell for Mn\textsubscript{18}GaSn\textsubscript{5}. (e), (f): Energy-dependent AHC curves near 1/6 Ga doping. The blue dashed lines indicate the results from supercell calculations. The red dashed line in (f) represents the weighted average AHC results of a larger 2×1×3 supercell under the same 1:5 ratio.
	}
	\label{fig4}
\end{figure}
The atomic numbers of Ge  and Ga differ by a single unit,  leading to more comparable atomic radii and SOC strengths. Following the validation of the VCA and non-VCA calculation for Ga-Sn doping, we further confirmed the Ga-Ge doping scenario exhibited similar consistency, as detailed in Appendix \ref{appB}.

Intrinsic AHC results for both Mn\textsubscript{3}Ga\textsubscript{1-x}Ge\textsubscript{x} and Mn\textsubscript{3}Ga\textsubscript{1-x}Sn\textsubscript{x} are calculated. Utilizing the VCA approach, we computed 21 distinct doping ratios, with $x$ varying from 0 to 1 in increments of 0.05.  As illustrated in Figs.~\ref{fig4}(a) and (b), the Fermi level experiences a gradual shift in response to changes in the doping ratio, coupled with a nuanced evolution in the overall curve morphology.  Furthermore, super cells of $1 \times 1 \times 2$ and $1 \times 1 \times 3$ are constructed in order to calculate the distinct ratios $x=\frac{1}{6}, \frac{1}{4},\frac{1}{3},\frac{1}{2},\frac{2}{3},\frac{3}{4},\frac{6}{6}$. The results of the VCA and supercell methods for both compounds diverge in value, yet the prevailing trend is quite analogous, as shown in Fig.~\ref{fig4}(c) and (d). To bolster the precision of our analysis, we implemented a Fermi-Dirac smoothing technique across all AHC curves, employing a parameter value of 0.003 eV. Therefore, the error bars represent the maximum and minimum AHC values within a 0.003 eV range above and below the Fermi level. As a result, when the AHC displays significant fluctuations in relation to the Fermi level, the corresponding error bars are appropriately expanded.

 In accordance with the predictions in preceding research~\cite{zhang2017strong,guo2019erratum}, a high AHC value is observed for both Mn\textsubscript{3}Ga\textsubscript{1-x}Ge\textsubscript{x} and Mn\textsubscript{3}Ga\textsubscript{1-x}Sn\textsubscript{x} at approximately x=5/6, which corresponds to 1/6 Ga doping(circled in Figs.~\ref{fig4} (c) and (d)). As depicted in Fig.~\ref{fig4}(d), in the compound of  Mn\textsubscript{3}Ga\textsubscript{1-x}Sn\textsubscript{x}, the peak is particularly sharp with its value of $700.27\mathrm{(\Omega \cdot cm)^{-1}}$, which is approximately 4.6 times larger than that of pure \ce{Mn_3Sn} (calculated $125.54\mathrm{(\Omega \cdot cm)^{-1}}$).  Furthermore, supercell calculations yield even higher values, reaching $865.31\mathrm{(\Omega \cdot cm)^{-1}}$(blue dotted line in Fig.~\ref{fig4}(f)). To further validate these results with respect to disorder effects, we also performed calculations on the optimally doped \ce{Mn48Ga2Sn10} on a larger $2\times1\times3$ supercell. Averaging over all unique, symmetry-inequivalent dopant configurations, we obtain a weighted average(weighted by the degeneracy of each configuration) anomalous Hall conductance (AHC) peak exceeding $750 [\mathrm{\Omega \cdot cm}]^{-1}$, as shown in the red dashed line in Fig. \ref{fig4} (f).

 In comparison, the curve for Mn\textsubscript{3}Ga\textsubscript{1-x}Ge\textsubscript{x} exhibited a relatively gentle change, forming a high AHC plateau, with values exceeding $450\mathrm{(\Omega \cdot cm)^{-1}}$ within a $x$ range from 0.55 to 0.85 (Fig.~\ref{fig4}(c)). The highest AHC value was found to occur at x=0.7, reaching $632.87\mathrm{(\Omega \cdot cm)^{-1}}$, which is approximately 3.4 times larger than that of pure \ce{Mn3Ge}. In this instance, a pronounced peak is observed for the 5:1 supercell calculation, which is illustrated by the blue dashed line in Figs.~\ref{fig4}(e), but its peak does not occur at the Fermi level with absolute precision. 

\subsection{Discussion}
In comparison with the conventional methods of regulating electron filling that involve the utilisation of excess or insufficient stoichiometric Mn, our strategy of in situ replacement of non-magnetic atoms attains a more extensive range of Fermi level  regulation. Concurrently, it circumvents the potential for excessive or insufficient Mn to induce phase transitions in the crystal or magnetic structure, thereby attaining a more substantial theoretical value (compared to Mn\textsubscript{2.4}Ga, 540$\mathrm{(\Omega \cdot cm)^{-1}}$ theoretically~\cite{song2024large}). Furthermore, these predictions, based on ideal structures, may represent conservative estimates, as structural relaxation effects like trimerization (as detailed in Appendix~\ref{appH}) were found to generally increase the AHC.

Experimentally, in the \ce{Mn3Z} (Z = Ga, Ge, Sn) materials, replacing Ge or Sn with Ga is expected to be structurally and chemically feasible. Firstly, \ce{Mn3Ga}, \ce{Mn3Ge} and \ce{Mn3Sn} share the same structural hexagonal lattice (space group: P6\textsubscript{3}/mmc), in which the Z atoms occupy equivalent 2c Wyckoff positions. This structural homogeneity eliminates the positional mismatch barrier of substitutional doping. Secondly, the minimal Pauling electronegativity differences ($\Delta E_N$ = 0.20 for Ga-Ge; $\Delta E_N$ = 0.15 for Ga-Sn~\cite{pauling1992nature}) suppresses the ionic bonding tendency and favours the formation of a metallic solid solution. It is noteworthy that the atomic radii mismatch between Ga and Ge ($\Delta r \approx 4\%$~\cite{slater1964atomic}) is well below the 15\% Hume-Rothery threshold~\cite{callister2020materials}, thereby enabling low-strain Ge site substitution. Furthermore, the larger size difference between Ga and Sn ($\Delta r \approx 10.4\%$) also serves to lower this limit, thereby suggesting that the risk of lattice distortion in Sn-based systems remains low.

\section{conclusion}
This study presents a significant exploration into the manipulation of the anomalous Hall effect  in Mn\textsubscript{3}Ga\textsubscript{1-x}Z\textsubscript{x} (Z = Ge, Sn) through selective doping. Utilizing first-principles density-functional theory calculations, we have demonstrated that by adjusting the electron filling through the substitution of Ga for Ge and Sn, we can significantly enhance the anomalous Hall conductivity. Key findings include the identification of specific Ga:Sn and Ga-Ge ratios that yield peak AHC values exceeding  $700\mathrm{(\Omega \cdot cm)^{-1}}$ and $630\mathrm{(\Omega \cdot cm)^{-1}}$, respectively.  The comparison of virtual crystal approximation (VCA) and supercell construction methods for doping revealed consistent trends, validating our computational approaches. Our analysis of the Berry curvature distribution provided insights into the mechanisms behind the AHE, with the discovery of BC hotspots that contribute  to the significant AHE.

In summary, our work offers a clear strategy for enhancing the AHE in non-collinear antiferromagnets through selective doping, which could be crucial for the development of advanced antiferromagnet-based technologies in data storage and processing. Future research will aim to experimentally validate these findings and explore further optimization of the AHE in these materials.

\section{Acknowledgments}

This work was supported by the National Key R\&D Program of China (Grant No. 2022YFA1402604), the National Natural Science Foundation of China(Grant No. T2394472), the Taishan Scholars Program, the International Collaboration Project (Grant No. B16001), the Fundamental Research Funds for the Central Universities of China (Grant No.KG16324501), the Key Research and Development Program of Shandong Province (Grant No. 2023CXPT055), and the Key Technology Research and Industrialization Program of Qingdao (Grant No. 24-1-2qljh-17-gx). We thank Dr. Yuhao Jiang from Beihang University for providing methodological guidance on the theoretical calculations in this study.
\section{Data Availability Statement}
The data that support the findings of this study are openly available in the Zenodo repository at https://doi.org/10.5281/zenodo.15469112~\cite{Wen2025_Dataset_Mn3Sn}, which includes Key input files for the DFT calculations and processed data required to generate each figure in this paper.

\appendix
\renewcommand{\thefigure}{A\arabic{figure}}
\setcounter{figure}{0}

\section{Magnetic momentum, energy, and magnetic stability}

\begin{figure}[htbp]
	\centering \includegraphics[width=8.6cm]{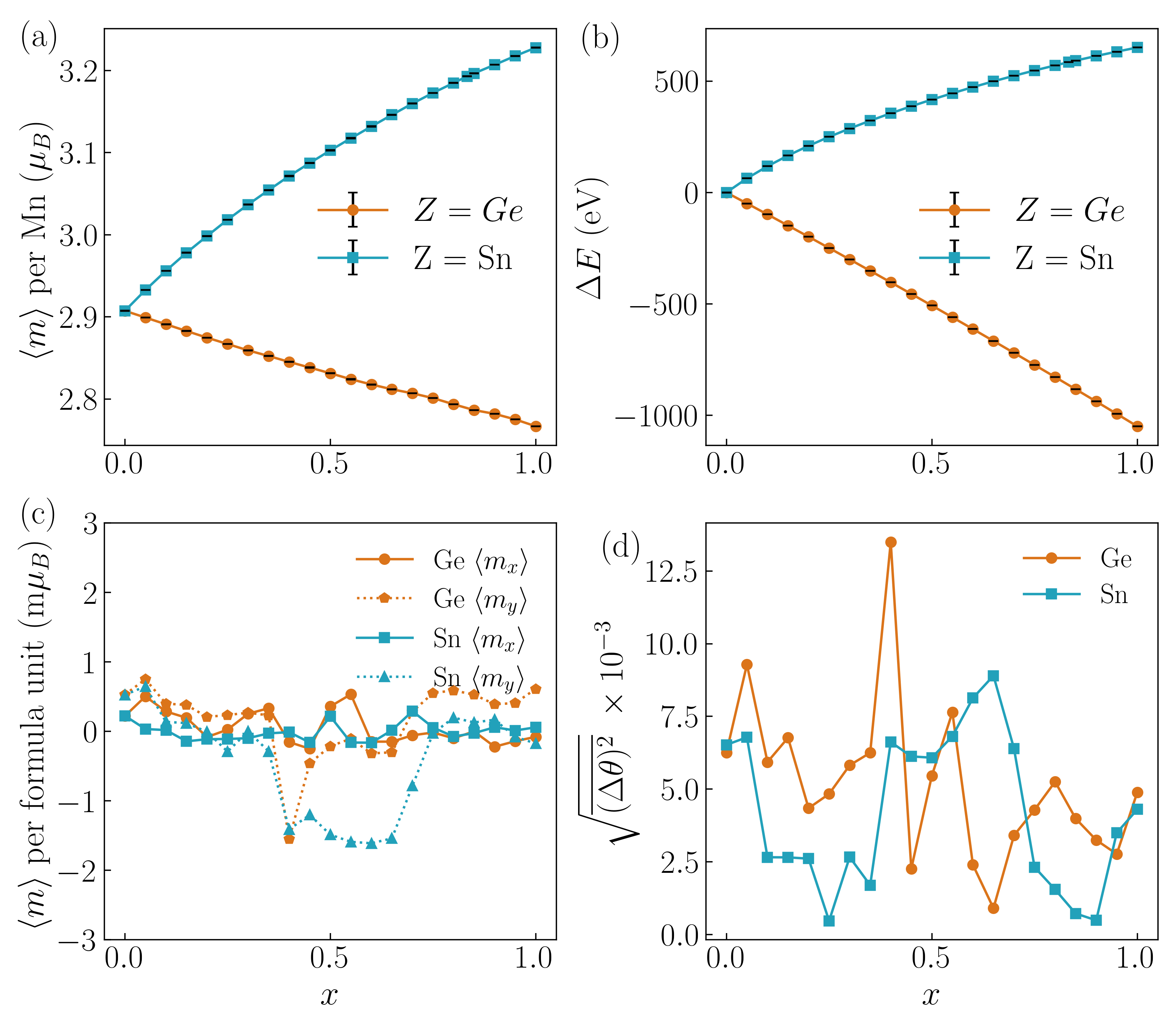} \caption{(a)  The doping ratio-dependent magnetic momentum of each Mn atom.  (b) Total energy change as a
	function of doping ratio. (c) Net in-plane magnetic moment per formula unit as a function of doping ratio x. (d) Root-mean-square angular deviation of Mn magnetic moments from the ideal $\Gamma_{1g}(Ay)$ configuration for both Ge- and Sn-based systems.}
	\label{figA1}
\end{figure}

The calculated atomic moment is approximately $\mathrm{3.23\mu_B/Mn}$ in \ce{Mn3Sn},  $\mathrm{2.91\mu_B/Mn}$ in \ce{Mn3Ga}, and $\mathrm{2.76\mu_B/Mn}$ in \ce{Mn3Ge}. These results are consistent with previous calculations~\cite{song2024large,guo2019erratum,kubler2014non} and are comparable to the experimental value~\cite{kren1970neutron,yamada1988magnetic,brown1990determination}. Furthermore,  change trend of total energy is similar, as shown in Fig.~\ref{figA1}(b). The total energy curve is convex upward, indicating a positive enthalpy of mixing. This suggests that non-equilibrium synthesis techniques should be employed in experimental realization.

We also analyzed the magnetic moments after the DFT self-consistent calculations. As illustrated in the Fig.~\ref{figA1}(c), the net in-plane magnetic moment per formula unit remains negligible, on the order of m$\mu_B$, across the entire doping range, showing a strong and robust Antiferromagnetism, which is also comparable to previous experimental findings~\cite{song2024large,chen2021anomalous}.
Furthermore, Fig~\ref{figA1}(d)  demonstrates that the root-mean-square deviation of the Mn magnetic moments from the ideal $\Gamma_{1g}(Ay)$ alignment is exceptionally small, on the order of $10^{-3}$ degrees. Crucially, we also confirmed that even when initializing the calculation with magnetic moments slightly perturbed from the ideal configuration, the system consistently relaxes back to this ground state, underscoring its energetic favorability and stability. 
%Nevertheless, the minute out-of-plane magnetic canting that gives rise to a weak ferromagnetism, as observed in earlier studies~\cite{li2023field,song2024large,chen2021anomalous}, was not discernible in our calculations. Fig.~\ref{figA1}(a) illustrates the gradual change of atomic moment in response to alterations in the Ga-Ge and Ga-Sn doping ratio. Furthermore,  change trend of total energy is similar, as shown in Fig.~\ref{figA1}(b).

\section{Half-doping case for Mn\textsubscript{3}Ga\textsubscript{0.5}Ge\textsubscript{0.5}}
\label{appB}
\begin{figure}[htbp]
	\centering \includegraphics[width=8.6cm]{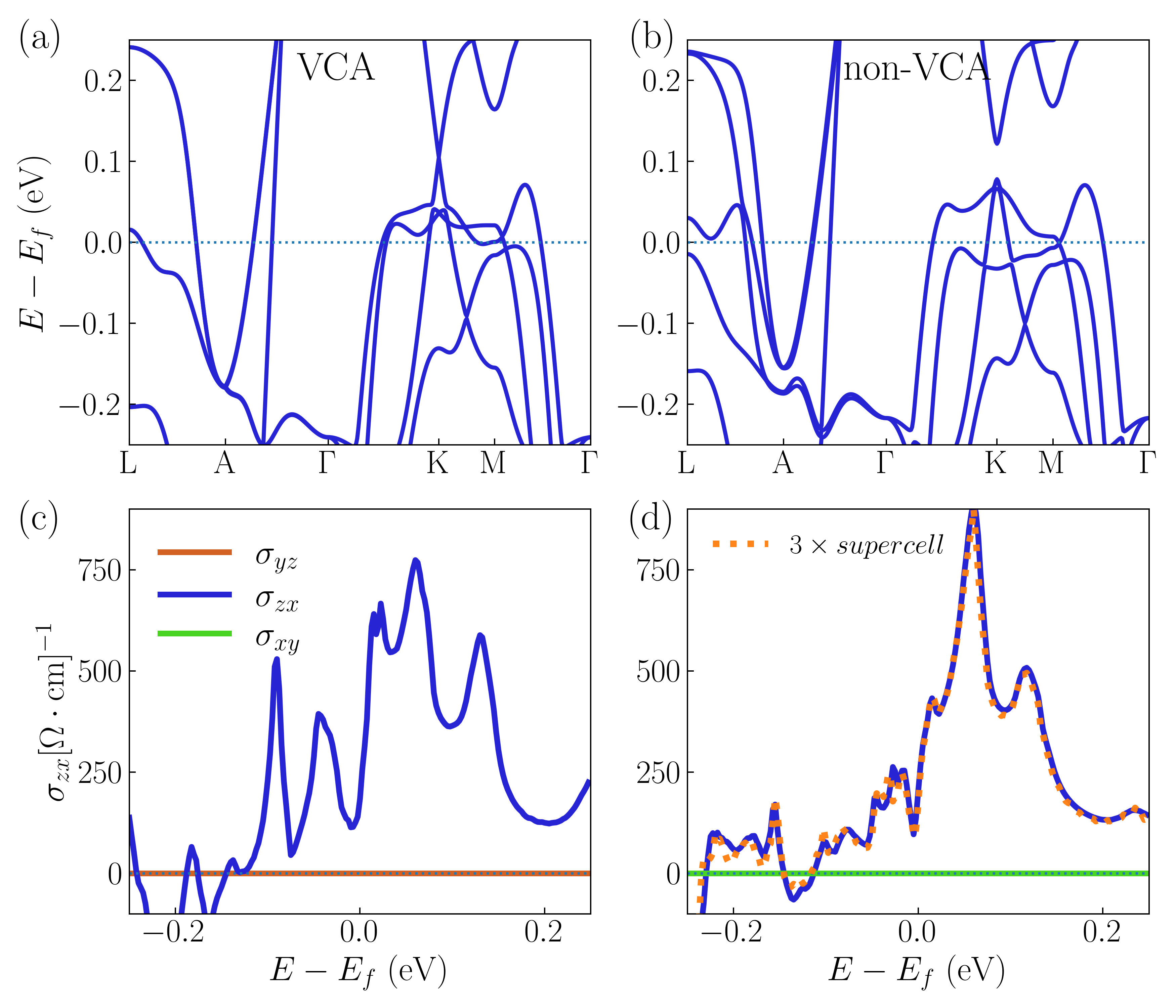} \caption{(a), (b) The electronic band structure for the VCA  and non-VCA scenarios, respectively. (c), (d) Their energy-dependent AHC.}
	\label{figA2}
\end{figure}
A comparison between a VCA and a non-VCA calculation is also conducted on case Mn\textsubscript{3}Ga\textsubscript{0.5}Ge\textsubscript{0.5}, following the same process as that described in Section III.A. As shown in Fig.~\ref{figA2}, the band structure exhibits substantial consistency. The primary distinction is the conspicuous emergence of an energy gap at approximately $0.05eV$ above $E_f$ at the K point in the non-VCA calculation. The AHC $\sigma_{zx}$ values obtained by these two calculations are $184.51\mathrm{(\Omega \cdot cm)^{-1}}$ and $203.20\mathrm{(\Omega \cdot cm)^{-1}}$ respectively. It should be noted that the $\sigma_{xy}$ and $\sigma_{zx}$ in the other two directions shown in Figs.~\ref{figA2}(c) and (d)  are strictly zero. This is due to the fact that the symmetry was manually set in the WannierBerri package when calculating the AHC.

\section{Ge-Sn doping}
\begin{figure}[htbp]
	\centering \includegraphics[width=8.6cm]{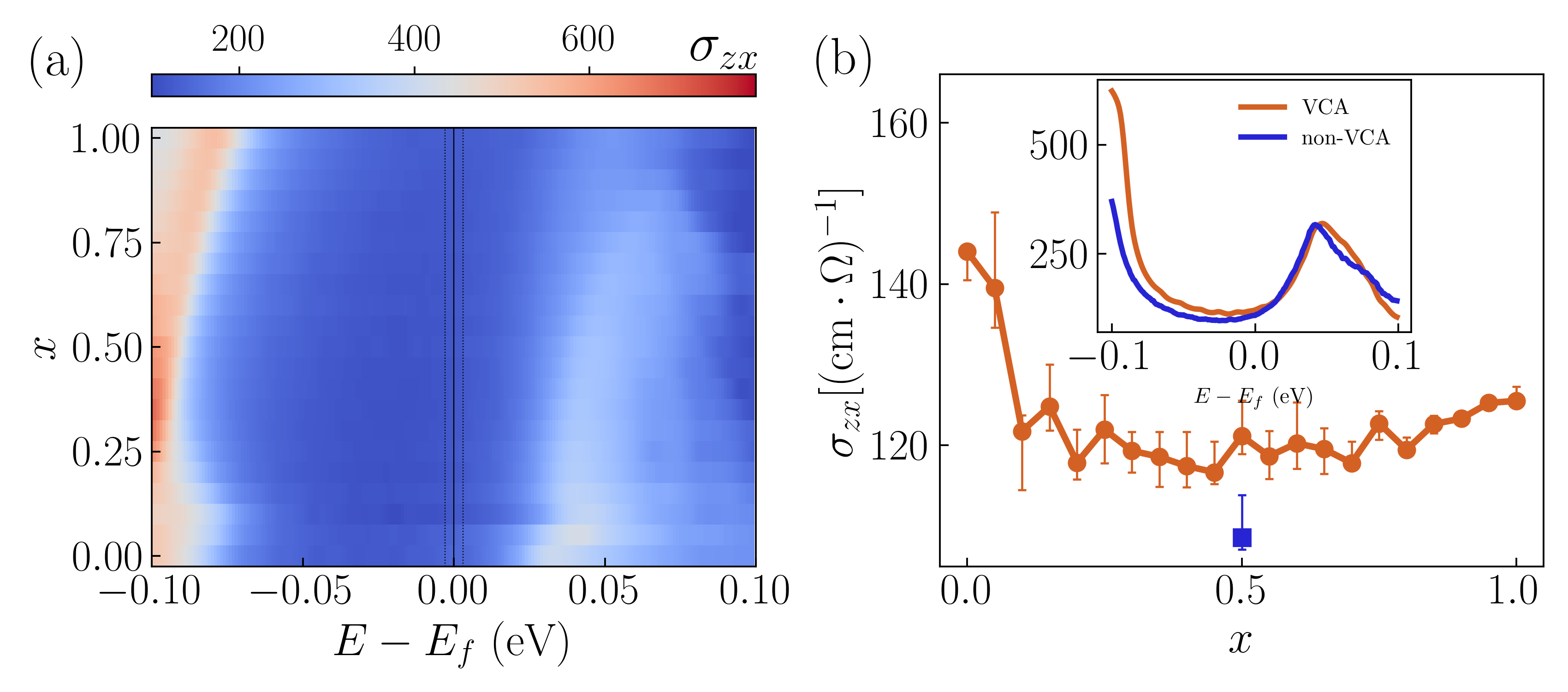} \caption{(a) Energy and doping ratio-dependent AHC $\sigma_{zx}$ pseudocolour plot of Mn\textsubscript{3}Ge\textsubscript{1-x}Sn\textsubscript{x}. The solid black line represents the neutral filling (Fermi level), while the dashed black line is employed to ascertain the error bar, which differs by 0.003 eV from the Fermi level. (b) AHC $\sigma_{zx}$ that undergoes a change in response to alterations in the doping ratio . The error bars represent the maximum and minimum values within the black dashed lines depicted in Fig. (a). The inset shows the  energy-dependent AHC curve for the half-doping case.}
	\label{figA3}
\end{figure}
Additionally, the Ge-Sn doping case was evaluated. Given that Ge and Sn are both members of the same main group and possess an identical number of valence electrons, a notable shift in the Fermi level is not anticipated. This is evidenced by the gradual evolution depicted in Fig.~\ref{figA3}(a). Consequently, the AHC curve depicted in Fig.~\ref{figA3}(b) exhibits less variation than that seen in Fig.~\ref{fig4}(c) and (d). It is noteworthy that the AHC calculated in this study is smaller than that reported in previous research~\cite{zhang2017strong}. This discrepancy may be attributed to the different setting of the trimerization of the lattice in those studies.

\section{Convergence test}
\begin{figure}[htbp]
	\centering \includegraphics[width=8.6cm]{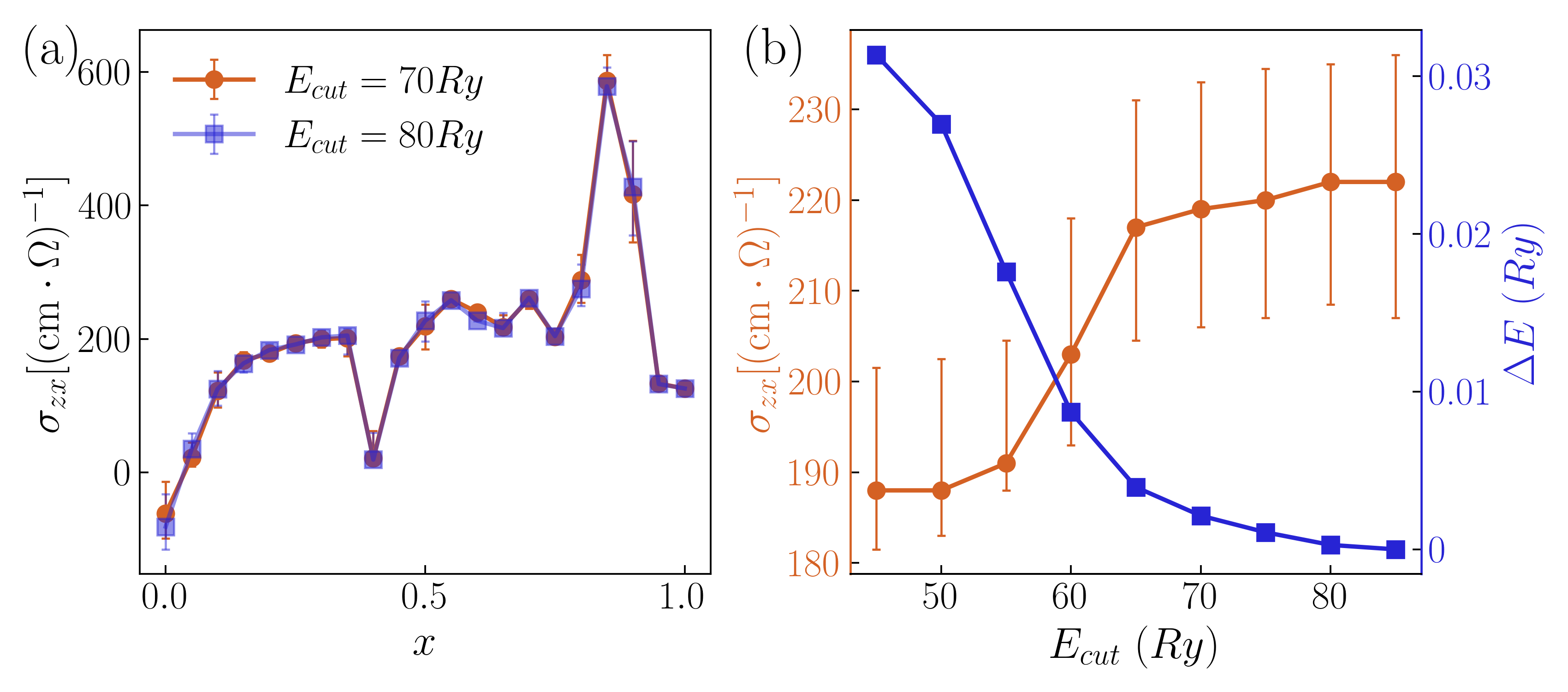} \caption{AHC of Mn\textsubscript{3}Ga\textsubscript{1-x}Sn\textsubscript{x} for different cut-off energies, as a function of doping ratio. (b) The AHC $\sigma_{zx}$ and total energy difference as a function of cut-off energy in the case of half-doping.}
	\label{figA4}
\end{figure}
The selected cutoff energy is 70 Ry (952.4 eV), which exceeds the recommended value indicated in the pseudopotential file but is consistent with the conventional parameters employed in earlier computational studies~\cite{chen2014anomalous}. The convergence of energy and AHC change was examined for a range of cut-off energy values in order to ascertain the results. As evidenced in Fig.~\ref{figA4}, the impact of employing an energy cutoff on AHC is less than the margin of error, and the observed trends remain consistent.

\section{Effect of trimerization}
\label{appE}
\begin{figure}[htbp]
	\centering \includegraphics[width=8.6cm]{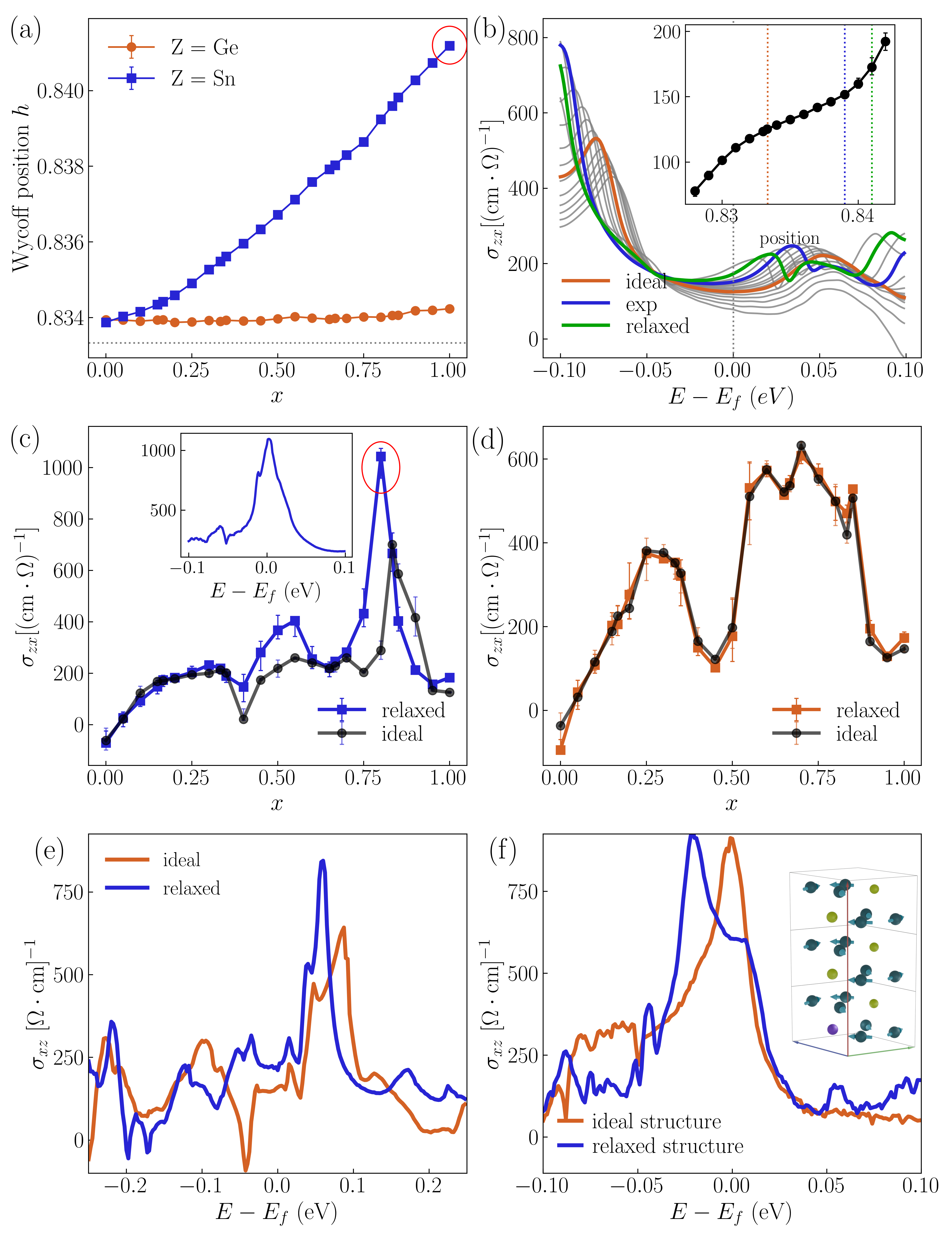} \caption{(a) Wyckoff positions after structural relaxation of Mn\textsubscript{3}Ga\textsubscript{1-x}Z\textsubscript{x}. Dashed line marks the ideal position. (b) Energy-dependent AHC curves of different Wyckoff positions. The ideal, experimental, and relaxed values are represented by the orange, blue, and green curves, respectively.The inset shows the variation of AHC as a function Wyckoff position. (c) doping ratio-dependent AHC curves of Mn\textsubscript{3}Ga\textsubscript{1-x}Sn\textsubscript{x} for different Wyckoff positions. The inset illustrates the peak observed at the coordinate x = 0.80 (encircled in red) when the Wyckoff position is relaxed. (d) doping ratio-dependent AHC curves of Mn\textsubscript{3}Ga\textsubscript{1-x}Ge\textsubscript{x} for different Wyckoff positions. (e) Comparison of the energy-dependent AHC for the ideal (orange) and relaxed (blue) structures in half-doped \ce{Mn6GaSn}. (f) A similar comparison for a 5:1 Sn:Ga supercell.}
	\label{figA5}
\end{figure}
The trimerization of the kagome plane in non-collinear antiferromagnets has the potential to influence its exchange interactions, which in turn can affect the AHC. A recent study has sought to elucidate the relationship between trimerization and the stabilisation of the helical phase~\cite{singh2020pressure}. Fig.~\ref{figA5} illustrates the impact of trimerization on AHC, based on VCA calculations for both Ga-Sn doping and Ga-Ge doping with varying degrees of trimerization. In an ideal scenario, the Wyckoff position for Mn 6h would be 5/6, with an increase in this value indicating enhanced trimerization of the kagome lattice. As illustrated in Fig.~\ref{figA5}(a), the greater radius of Sn atoms relative to Ga and Ge results in a more pronounced trimerization degree and a greater deviation from the ideal Wyckoff position. The calculated value is also higher than the experimental value. Fig.~\ref{figA5}(b) provides a detailed picture of the AHC calculation results for \ce{Mn3Sn} at varying Wyckoff positions. The AHC value demonstrates an increase with the advancement of trimerization, a trend that is also observed in \ce{Mn3Ge}. In comparison with the AHC curve of Mn\textsubscript{3}Ga\textsubscript{1-x}Ge\textsubscript{x} (see fig.~\ref{figA5}(d)), the doping ratio-dependent AHC curve of Mn\textsubscript{3}Ga\textsubscript{1-x}Sn\textsubscript{x} (fig.~\ref{figA5}(c)) at relaxed positions exhibits a greater deviation from ideal positions, with a peak at x=0.80 and a value exceeding $1000\mathrm{(\Omega \cdot cm)^{-1}}$. This is primarily attributable to the differences in Wyckoff position as previously discussed.

Furthermore, additional calculations were performed for non-VCA structures, with full atomic relaxation.  Our calculations demonstrate that, in semi-doped systems such as $\ce{Mn6GaSn}$, the presence of different atomic species (Ga vs. Sn) in adjacent layers can result in an asymmetric distortion of the Mn kagome triangle. The calculations have been presented in Fig~\ref{figA5}(e) for the half-doped case and Fig~\ref{figA5}(f) for a 5:1 Sn:Ga case. These compare the AHC for the ideal (unrelaxed) structures with those that have been relaxed. The results consistently demonstrate that structural relaxation significantly enhances the peak AHC values. It is merely the case that the peak position may be different. This finding indicates that the AHC values calculated from ideal structures in the main text may be considered conservative estimates, suggesting that the achievable AHC could be larger.

\section{Other layers of BC hotspots}
\label{appF}
\begin{figure}[htbp]
	\centering \includegraphics[width=8.6cm]{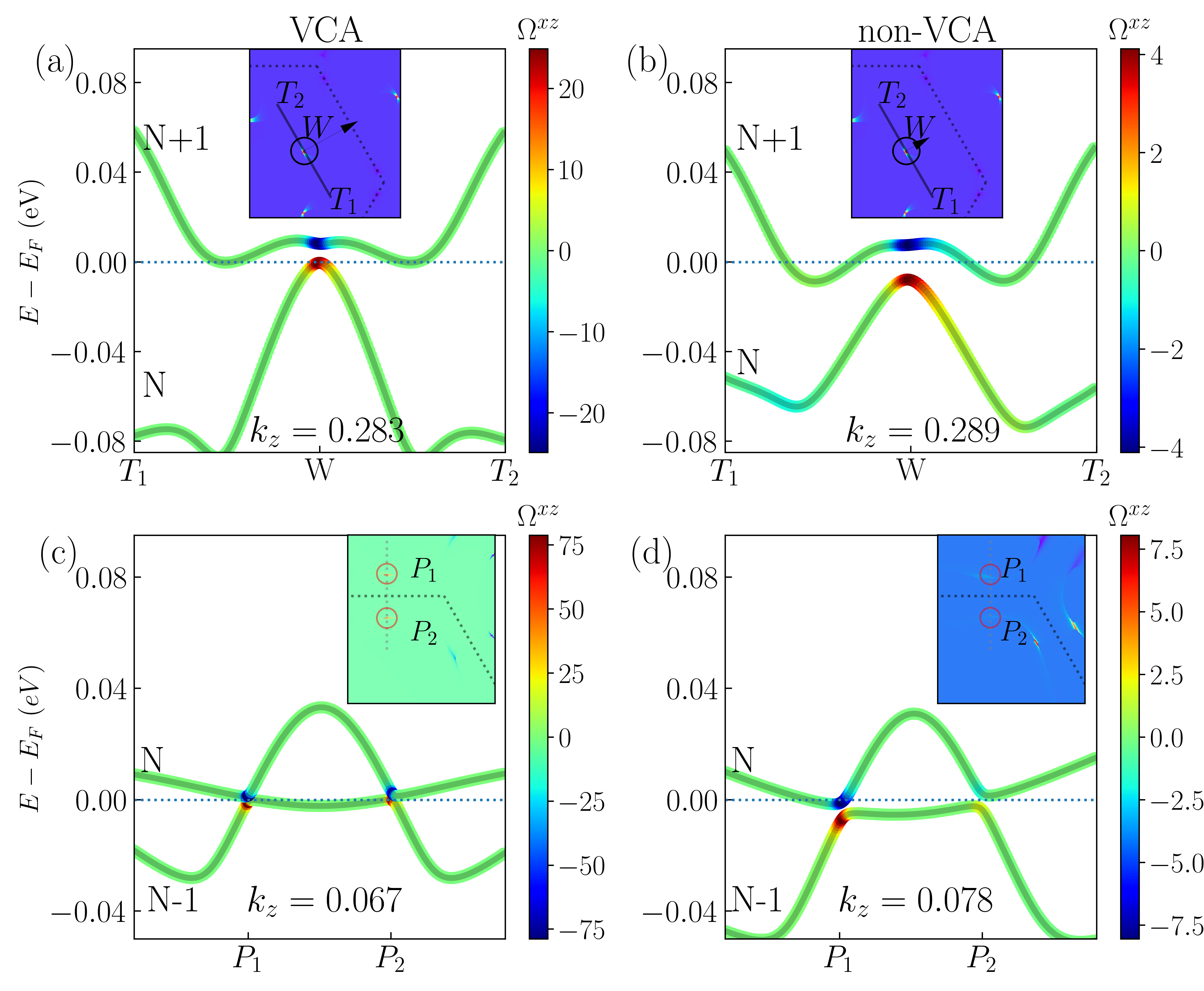} \caption{(a) and (b): hotspots of purple plane in fig.~\ref{fig3}(a) and (b), respectively, as well as fat bands traversing them. (c) and (d): hotspots of blue plane in fig.~\ref{fig3}(a) and (b),respectively, as well as fat bands traversing them.}
	\label{figA6}
\end{figure}
In  Fig.~\ref{figA6}, we present additional calculations of Berry curvature  hotspots on other layers of the half-doping case(i.e. the purple and blue planes depicted in Figs.~\ref{fig3}(a) and (b)), as coordinate $k_z$  are written in the plots. for both VCA and non-VCA calculations, Fermi level crosses two gaps around the hotspots, and both have similar band shape. On the k-path in the same direction, the fat band calculated by VCA approach has perfect symmetry(see Figs.~\ref{figA6} (a) and (c)), whereas the symmetry of the fat band calculated by the non-VCA approach is slightly disturbed(see Figs.~\ref{figA6} (b) and (d)). This finding aligns with our previous analysis. It is noteworthy that in the three-layer system, the minor gaps traversed by $E_f$ are constituted by disparate energy bands, the indices of which are indicated in the plots. 

\section{Berry Curvature Analysis and AHC Enhancement Mechanism}
\begin{figure}[htbp]
	\centering \includegraphics[width=8.6cm]{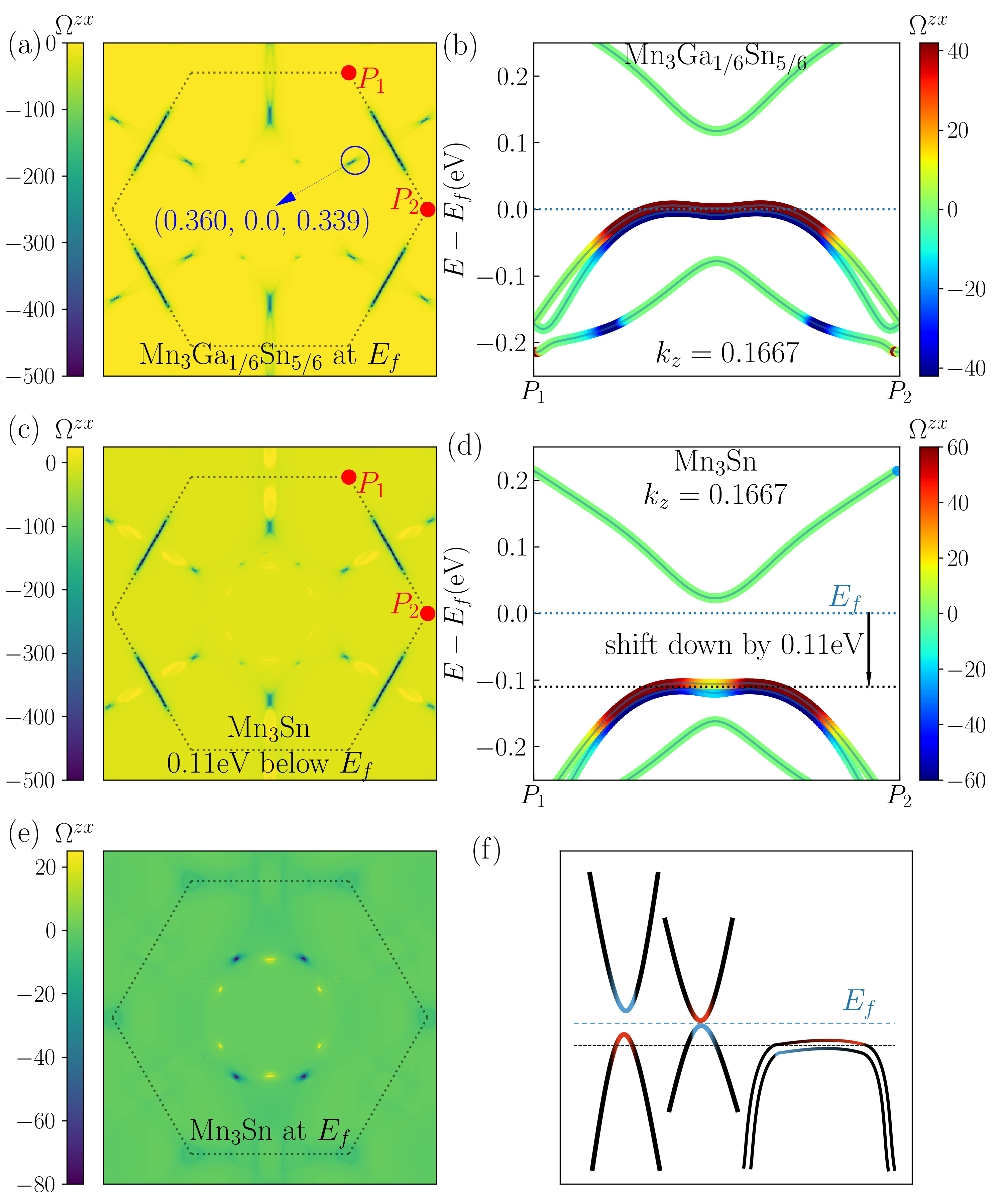} \caption{(a) Berry curvature  $\Omega^{xz}$ for optimally doped $\mathrm{Mn3Ga_{1/6}Sn{5/6}}$  projected onto the $k_x-k_y$ plane. (b)  Berry curvature $\Omega^{xz}$ for  $\mathrm{Mn3Ga_{1/6}Sn{5/6}}$ fat band along $P_1-P_2$ path at fixed $k_z = 0.1667$. (c) Projected $\Omega^{xz}$ for pure \ce{Mn3Sn} with its $E_f$ artificially shifted down by 0.11 eV. (d) Fatband for pure \ce{Mn3Sn} along the same $P_1-P_2$ path. (e) Projected $\Omega^{xz}$ for pure \ce{Mn3Sn}. (f) Schematic diagram illustrating how hole doping shifts the Fermi level downwards to intersect the large scale, pre-existing Berry curvature peak of the parent compound.}
	\label{figA7}
\end{figure}
Furthermore, the distributions of the Berry curvature of Mn\textsubscript{3}Ga\textsubscript{1/6}Sn\textsubscript{5/6} were examined to ascertain the reason for the pronounced peak observed. The Berry curvature components of $\Omega^{zx}$ were projected onto the kx-ky plane by summing them along kz. Fig.~\ref{figA7}(a) depicts the projected Berry curvatures of Mn\textsubscript{3}Ga\textsubscript{1/6}Sn\textsubscript{5/6} with the Fermi level situated at the charge neutral point. Two distinct layers of Berry curvature hotspots are evident. The upper  is observed at $kz=0.339$ and exhibits a similar characteristic to that of half-doping case, including the same orientation and same symmetry of distribution. The lower layer, however, occurs at the boundary of the First Brillouin zone, and the hot spots extend to form 'hotlines'. Its fat band is plotted in Fig. ~\ref{figA7}(b), with the k-path connecting two corners of the First BZ, as marked by red dots in Fig.~\ref{figA7}(b). From this, one can readily identify the origin of the significant large anomalous Hall conductivity (AHC): a larger k-space range where the energy gap between occupied bands and the empty bands are small. It is noteworthy that the number of valence electrons here is 49.6667, and the two states bands crossing Fermi level shown in the figure are the 49th and 50th energy bands, respectively.

To clarify the mechanism behind the AHC enhancement, we performed a comparative analysis shown in Fig. A7(c-f). Our analysis confirms that the enhancement is predominantly driven by a rigid-band-like shift of the Fermi level($E_f$).  The parent compound \ce{Mn3Sn} already possesses large intrinsic Berry curvature hotspots that result in an AHC peak located approximately 0.11 eV below its natural Fermi level. As Ga is a p-type dopant relative to Sn, its substitution introduces holes and lowers the Fermi level to align with these pre-existing hotspots, as schematically illustrated in Fig. ~\ref{figA7}(f). 
The distribution for the optimally doped $\mathrm{Mn_3Ga_{1/6}Sn_{5/6}}$ at its natural $E_f$(Fig. ~\ref{figA7}(a))is remarkably similar to that of pure \ce{Mn3Sn} when its $E_f$ is artificially lowered by 0.11 eV (Fig. A7(c)). 
The distribution for pure \ce{Mn3Sn} at its natural E is shown in Fig.~\ref{figA7}(e) for contrast.Furthermore, the Berry-curvature-weighted band structures ('fatbands') along an identical k-path are nearly identical for the optimally doped case (Fig.~\ref{figA7}(b)) and the shifted pure case (Fig. ~\ref{figA7}(d)), confirming the validity of the Fermi level tuning mechanism.

\section{Magnetic structures of the Mn}
\label{appH}
\begin{figure}[htbp]
	\centering \includegraphics[width=8.6cm]{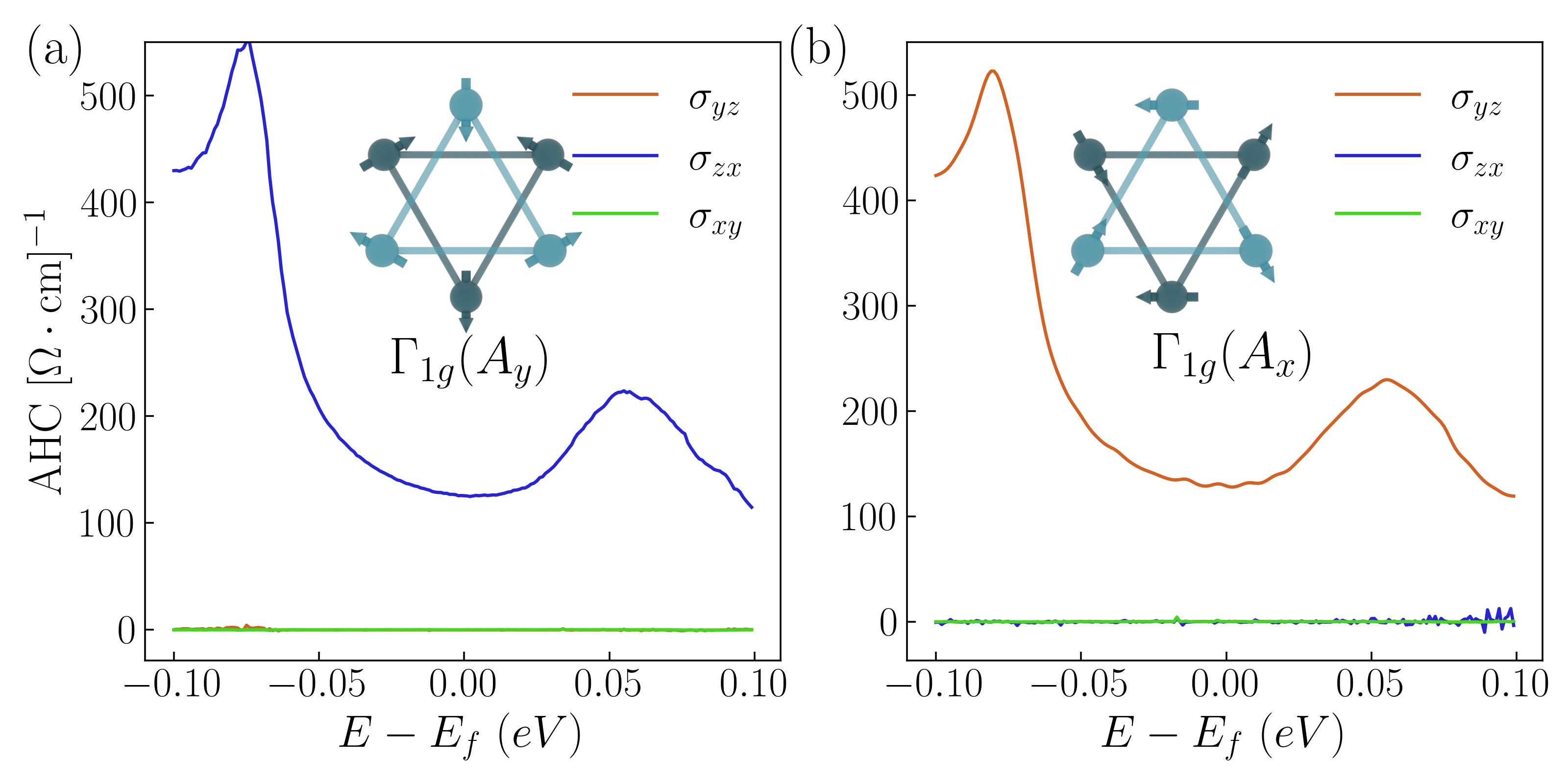} \caption{The Energy-AHC curves of different magnetic order. The order of the Mn atoms are shown in the insets. }
	\label{figA8}
\end{figure}
 
The AHC was calculated for all four magnetic orders of \ce{Mn3Sn}~\cite{soh2020ground}, and only the two E1g magnetic orders yielded non-zero AHC, which is consistent with the results of the symmetry analysis. The two E1g magnetic orders have  nonsymmorphic symmetry $\{ M_x|\tau =2/c \}$ and  $\{ M_y|\tau =2/c \}$, which results in their AHC being non-zero only in the y-z and z-x directions. The results of these two orders are almost identical, as illustrated in Fig.~\ref{figA8}. %less E by 0.001Ry

\section{Effect of Smearing}
\label{appI}
\begin{figure}[htbp]
	\centering \includegraphics[width=8.6cm]{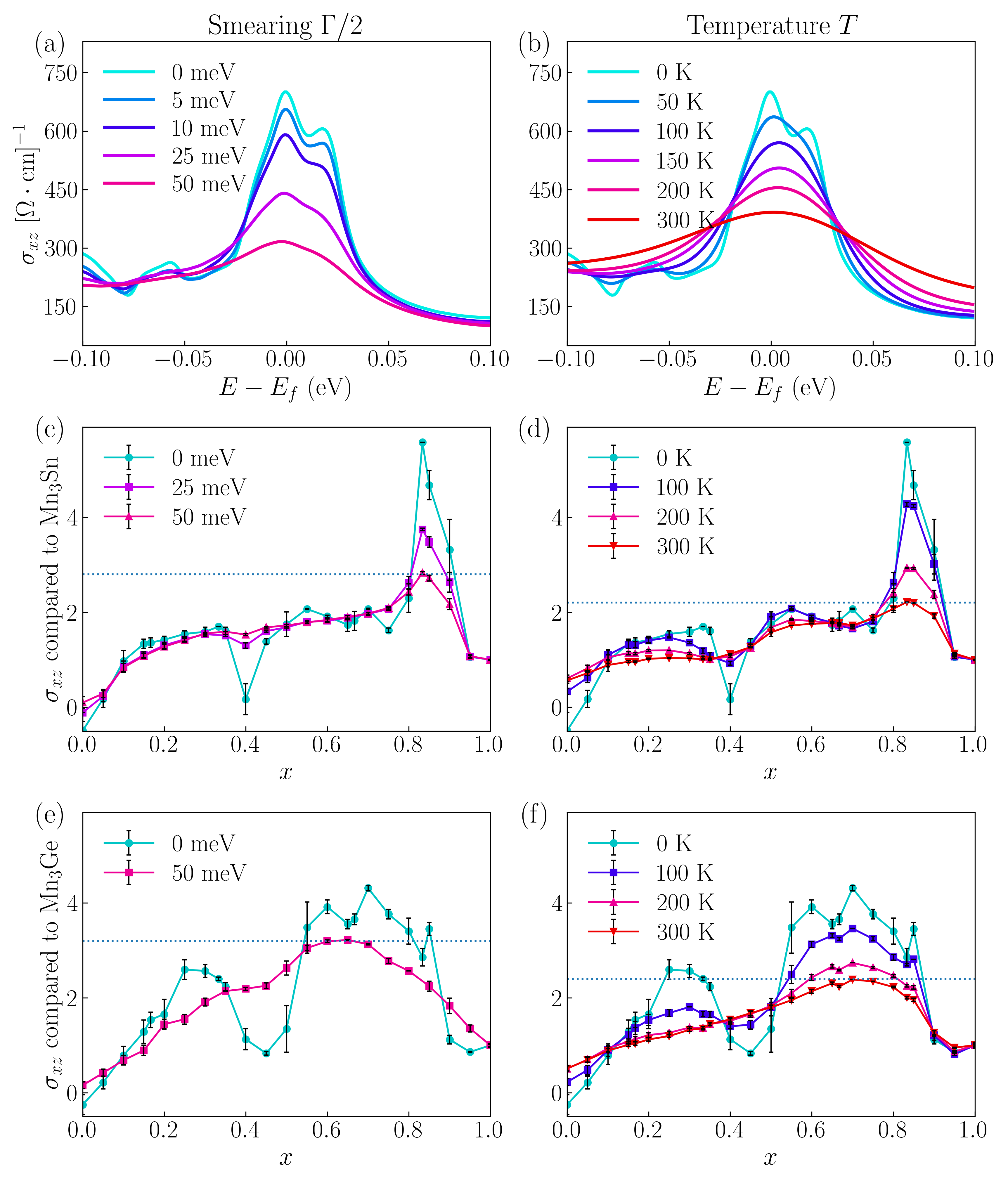}
	 \caption{Effect of finite lifetime broadening ($\Gamma$) and temperature ($T$) on the anomalous Hall conductivity (AHC). 
	 	(a) Energy-dependent AHC ($\sigma_{xz}$) for the peak composition Mn\textsubscript{3}Ga\textsubscript{1/6}Sn\textsubscript{5/6} calculated with different Lorentzian broadening parameters $\Gamma/2$.
	 	(b) Energy-dependent AHC for the same composition at various temperatures, showing thermal smearing.
	 	(c) and (e) AHC enhancement factor (relative to pure Mn\textsubscript{3}Sn and Mn\textsubscript{3}Ge, respectively) as a function of Ga concentration $x$ for several broadening values.
	 	(d) and (f) AHC enhancement factor as a function of Ga concentration $x$ at different temperatures.}
	\label{figA9}
\end{figure}

In any real material, the electronic states have a finite lifetime due to scattering from impurities, defects, and phonons. Furthermore, at any non-zero temperature, the electron occupations are thermally broadened. Both of these effects can influence the measured Anomalous Hall Conductivity (AHC), especially when sharp features occur.

%First, we consider the finite lifetime of electronic states, which introduces a Lorentzian broadening in the energy domain, with a full width at half maximum (FWHM) of $\Gamma = \hbar/\tau$, where $\tau$ is the scattering time. 
We first modeled the effect of a finite electronic lifetime (due to scattering) by applying a Lorentzian broadening term, $\Gamma$, to the Berry curvature calculation, 
which is modeled by introducing a complex broadening term, $i\Gamma/2$, into the energy denominator of the Berry curvature expression Eq.\ref{eq2}. 
As shown in Fig.~\ref{figA9}(a), for the optimally doped Mn\textsubscript{3}Ga\textsubscript{1/6}Sn\textsubscript{5/6} this broadening suppresses and widens the AHC peak, as expected. However, even with a substantial broadening of $\Gamma /2 $ = 50 meV(corresponding to a realistic scattering time of approximately $1.32 \times 10^{-14}$s), the key conclusion that Ga doping enhances the AHC holds firm. 
%As shown in Fig.~\ref{figA9}(a) for the optimally doped Mn\textsubscript{3}Ga\textsubscript{1/6}Sn\textsubscript{5/6} composition, the sharp AHC peak calculated at the clean limit ($\Gamma=0$) is suppressed and broadened as the smearing parameter $\Gamma$ increases, which is an expected outcome. However, even with a substantial broadening of $\Gamma/2 = 50$\,meV (corresponding to a realistic scattering time of approximately $1.32 \times 10^{-14}$s), 
 More importantly, the overall trend of AHC enhancement as a function of Ga doping holds firm. As shown in Figs.~\ref{figA9}(c) and (e), with $\Gamma/2 = 50$\,meV, a significant AHC enhancement is still predicted. The peak AHC value in the Ga-doped Sn system is still $\sim$2.8 times that of pure Mn\textsubscript{3}Sn, and the peak in the Ga-doped Ge system is $\sim$3.2 times that of pure Mn\textsubscript{3}Ge.
 
Next, we analyzed the effect of temperature. The underlying magnetic order is stable well above 300 K, as evidenced by the high Néel temperatures of the parent compounds~\cite{balk2019comparing,cable1993magnetic,kren1970neutron}.
%Second, we analyze the effect of thermal disturbances. The stability of the non-collinear magnetic order, a prerequisite for a robust AHE, is well-supported by the high Néel temperatures of the parent compounds ($T_N \approx 420$\,K for Mn\textsubscript{3}Sn, $395$\,K for Mn\textsubscript{3}Ge, and $460$\,K for Mn\textsubscript{3}Ga), all of which are well above room temperature.
 %This indicates the essential magnetic structure should not be significantly disrupted by thermal fluctuations at 300 K. While a full first-principles calculation of all thermal effects is computationally prohibitive, 
 We approximated the effect of thermal smearing by incorporating the finite-Temperature Fermi-Dirac distribution into our calculations into the Kubo formula in Eq.\ref{eq1}. 
The results of this analysis are shown in Figs.~\ref{figA9}(b), (d), and (f). For the optimally doped Mn\textsubscript{3}Ga\textsubscript{1/6}Sn\textsubscript{5/6}, the AHC peak broadens and its magnitude decreases as temperature increases. The doping-dependent AHC curves at several temperatures confirm that while the peak AHC values are reduced from their 0 K values, but significant enhancement remains at room temperature. Specifically, at 300 K, the predicted AHC for optimally doped Mn\textsubscript{3}Ga\textsubscript{x}Sn\textsubscript{1-x} is still enhanced by a factor of approximately 2.2 relative to pure Mn\textsubscript{3}Sn, and for Mn\textsubscript{3}Ga\textsubscript{x}Ge\textsubscript{1-x}, the enhancement factor is approximately 2.4.

%In summary, although scattering and thermal effects quantitatively reduce the AHC values from the ideal theoretical limit, our central finding—that Ga doping is a potent strategy for significantly enhancing the AHC in these materials—remains robust and should be experimentally observable at room temperature.
Therefore, although scattering and thermal effects quantitatively reduce the AHC from the ideal 0 K limit, our central finding—that Ga doping is a potent strategy for enhancing the AHC—remains robust and should be observable in experiments at room temperature.

%\bibliography{Mn3X_ref_abb}% Produces the bibliography via BibTeX.

\end{document}